# Superconductivity in 3R-$Ta_{1-x}M_xSe_2$ (M = W, Mo)

Huixia Luo*, Weiwei Xie, Elizabeth M. Seibel and R. J. Cava*

Department of Chemistry, Princeton University, Princeton, New Jersey 08544, USA.

* rcava@princeton.edu; huixial@princeton.edu

**Abstract**

The 3-layer (3R) polytype of $TaSe_{2-x}Te_x$ shows a superconducting transition temperature that is between 6 and 17 times higher than that of the two-layer (2H) polytype. The remarkable difference in $T_c$, although clearly associated with a difference in polytype, could have been due to an electronic effect specific to the Te-Se substitution. Here we report that small amounts of Mo or W doping lead to a 2H to 3R polytype transition in $Ta_{1-x}Mo_xSe_2$ and $Ta_{1-x}W_xSe_2$. The 3R polytype materials are again found to have substantially higher $T_c$s (~ 2 K for $Ta_{0.9}W_{0.1}Se_2$ and $Ta_{0.9}Mo_{0.1}Se_2$) than the 2H material (0.15 K). This eliminates the possibility that any special characteristics of the Te/Se substitution are responsible for the dramatic difference in $T_c$, and instead shows that a three-layer stacking sequence is strongly preferred for superconductivity over a two-layer stacking sequence in the $TaSe_2$ system.

## 1. **Introduction**

The layered $MX_2$ transition-metal dichalcogenides (TMDs), (M = Mo, W, V, Nb, Ta, Ti, Zr, Hf, and Re, and X = Se, S or Te), have been studied for decades due to the electronic properties such as superconductivity, large thermoelectric effect and anomalous magnetoresistance that emerge due to their low dimensionality[1-5]. Structurally, TMDs have (2D) X-M-X layers, with M in trigonal prismatic or octahedral coordination with X, and van der Waals gaps. The different possible M-X coordinations and different stacking sequences of the $MX_2$ layers lead to structural polymorphism, which enriches their electronic properties.[6,7] $MoTe_2$, for example, has two kinds of polymorphs, a 2H-form ($\alpha$-$MoTe_2$)[8] and a distorted 1T form ($\beta$-$MoTe_2$)[9]; $\alpha$-$MoTe_2$ shows semiconducting behavior and $\beta$- $MoTe_2$ is metallic.[10,11] Recently, four different polytypes of $Mo_{1-x}Nb_xTe_2$ have been reported to display different electronic properties. [12]

$TaSe_2$ is one of the highly polymorphic and polytypic TMDCs.[13] In both its 2H and 3R polytypes (Figure 1a), Ta is found in trigonal prismatic coordination in Se-Ta-Se layers that are stacked along the *c* axis of hexagonal (2H) or rhombohedral (3R) cells; the structure repeats after 2 layers in the 2H form and 3 layers in the 3R form[14-16]. The polytype commonly obtained is 2H; the 3R form can be synthesized, but it is not the stable variant. Recently, we successfully prepared stable 2H, 3R, 1T, and monoclinic layered materials in the $TaSe_{2-x}Te_x$ system; the 3R polytype shows a superconducting transition temperature that is between 6 and 17 times higher than that of 2H-$TaSe_2$, where $T_c$ is 0.15 K.[17]

Here we report that the 3R structure is also stabilized in $TaSe_2$ by substitution of Mo or W for Ta in $Ta_{1-x}M_xSe_2$ at the 10 % level. We also report that superconductivity again emerges at relatively high temperatures in these 3R-polytypes. The highest superconducting transition temperatures ($T_c$s) are ~ 2 K for both $Ta_{0.9}W_{0.1}Se_2$ and $Ta_{0.9}Mo_{0.1}Se_2$, which is close to what is seen for 3R-$TaSe_{1.65}Te_{0.35}$. We conclude that it is the three-layer sequence of 3R-$TaSe_2$ that leads to the dramatic increase in $T_c$ compared to two-layer 2H-$TaSe_2$, not any special characteristic of the substituting elements themselves.

## 2. **Experimental**

Polycrystalline samples of $Ta_{1-x}M_xSe_2$ (M = W, Mo; $0.07 \leq x \leq 0.2$) were made by solid state reaction in evacuated silica tubes. First, mixtures of high-purity fine powders of Ta (99.8%), Mo (99.95%) or W (99.95%) and Se (99.999%) in the appropriate stoichiometric ratios were thoroughly ground, pelletized and heated in sealed evacuated silica tubes at a rate of 1 $^{o}$C/min to 800 $^{o}$C and held there for 120 h. Subsequently, the as-prepared powders were reground, re-pelletized and sintered again, heated at a rate of 3 $^{o}$C/min to 800 $^{o}$C and held there for 120 h.

The identity and phase purity of the samples was determined by powder X-ray diffraction (PXRD) using a Bruker D8 ECO with Cu Kα radiation and a Lynxeye detector. Le Bail fits were performed on the powder diffraction data with the use of the FULLPROF diffraction suite using Thompson-Cox-Hastings pseudo-Voigt peak shapes.[18] Measurements of the temperature dependence of the electrical resistivity and heat capacity were performed in a Quantum Design Physical Property Measurement System (PPMS). For the superconducting samples, $T_c$ was taken as the intersection of the extrapolation of the steepest slope of the resistivity ρ(T) in the superconducting transition region and the extrapolation of the normal state resistivity ($\rho_n$).[19] Heat capacities for $Ta_{0.9}W_{0.1}Se_2$ and $Ta_{0.9}Mo_{0.1}Se_2$ were measured in the PPMS equipped with a $^3$He cryostat.

The electronic structures of the hypothetical model compounds "$Ta_{8/9}Mo_{1/9}Se_2$" and "$Ta_{8/9}W_{1/9}Se_2$", which simulate the superconducting materials, were calculated using the WIEN2k code, employing the full-potential linearized augmented plane wave method (FP-APW) with local orbitals. Electron correlation was treated by the generalized gradient approximation.[20] For valence states, relativistic effects were included through a scalar relativistic treatment, and core states were treated fully relativistically. To simulate the effect of the Mo and W doping, a 3 × 3 supercell in the *ab*-plane, in the experimental space group and with the experimentally determined unit cell and structure, was used in the calculations. The supercell content, with Mo/W occupying 3*a* (0, 0, 0) site, was $Ta_{24}(Mo/W)_3Se_{54}$; this simulates, through a crystallographically ordered cell (although the actual substitution is random) the effect of 11 % Mo or W doping of $TaSe_2$; such models give a good approximation to the actual case for small amounts of random doping.

3. **Results and Discussion**

Figure 1 shows the crystal structures of the 2H and 3R polytypes of $TaSe_2$, which consist of edge-shared trigonal prisms of $MX_6$, with 2 and 3 layers, respectively, in the repeating cell along *c*. When doping with Mo or W, 2H-$TaSe_2$ transforms to a non-centrosymmetric rhombohedral 3R structure (*space group R3m*). This is evidenced by the powder X-ray diffraction (PXRD) patterns: the PXRD patterns for 3R-$Ta_{0.9}W_{0.1}Se_2$ and 3R-$Ta_{0.9}Mo_{0.1}Se_2$ are shown in the main panel of Figure 1b and c, respectively.

Figure 2 shows the temperature dependence of the normalized electrical resistivity, ($\rho/\rho_{300K}$), for the polycrystalline 3R-$Ta_{1-x}W_xSe_2$ ($0.07 \leq x \leq 0.125$) and 3R-$Ta_{1-x}Mo_xSe_2$ ($0.07 \leq x \leq 0.2$) samples. All the 3R samples have resistivities below 10 mΩ-cm at 300 K, which is similar to what is observed for the 3R-$TaSe_{2-x}Te_x$ compounds.[17] The residual-resistivity-ratio is very small, RRR = $\rho_{300K}/\rho_n < 1.5$, or otherwise the samples are slightly semiconducting, which we interpret as a reflection of the substantial Ta-W or (Ta/Mo) disorder present. At low temperatures, a clear, sharp ($\Delta T_c < 0.2$ K) drop of ρ(T) is observed, signifying the onset of superconductivity (inset of Figure 2a). For both materials, $x = 0.1$ shows the highest $T_c$, ~ 2 K.

Further information on the electronic properties and superconductivity in 3R-$Ta_{0.9}W_{0.1}Se_2$ and 3R-$Ta_{0.9}Mo_{0.1}Se_2$, was obtained from the heat capacity measurements. The main panels of Figure 3a and b show the temperature dependence of the specific heat, $C_p/T$, versus $T^2$, for 3R-$Ta_{0.9}W_{0.1}Se_2$ and 3R-$Ta_{0.9}Mo_{0.1}Se_2$, in 0 field and $\mu_0H = 5T$ applied field. The normal state specific heat at low temperatures (but above $T_c$) obeys the relation of $C_p/T = \gamma + \beta T^2$, where γ is the Summerfield electronic specific heat coefficient and β is the low temperature limit of the lattice heat capacity, the latter of which is a measure of the Debye Temperature ($\theta_D$). We obtained the electronic specific heat coefficients $\gamma = 7.26$ mJ·mol$^{-1}$·K$^{-2}$ for 3R-$Ta_{0.9}W_{0.1}Se_2$ and $\gamma = 5.81$ mJ·mol$^{-1}$·K$^{-2}$ for 3R-$Ta_{0.9}Mo_{0.1}Se_2$, and the phonon specific heat coefficients $\beta = 0.68$ mJ·mol$^{-1}$·K$^{-4}$ for 3R-$Ta_{0.9}W_{0.1}Se_2$ and $\beta = 0.75$ mJ·mol$^{-1}$·K$^{-4}$ for 3R-$Ta_{0.9}Mo_{0.1}Se_2$, based on fitting the data in the temperature range of 2 - 10 K. Using these values of β, we estimate the Debye temperatures by the relation $\theta_D = (12\pi^4 nR/5\beta)^{1/3}$, where n is the number of atoms per formula unit (n = 3), and R is the gas constant; the $\theta_D$ values are found to be ~203 K for 3R-$Ta_{0.9}W_{0.1}Se_2$ and ~198 K for 3R-$Ta_{0.9}Mo_{0.1}Se_2$, which are within error of each other. As shown in the insets for Figures 3a and b, both materials display a large specific heat jump at $T_c$. The superconducting transition temperatures are in excellent agreement with the $T_c$s determined in the ρ(T) measurements. Using the equal area method, we estimate $\Delta C/T_c$ = 9.5 mJ $mol^{-1}$ $K^{-2}$ for 3R-$Ta_{0.9}W_{0.1}Se_2$ and $\Delta C/T_c$ = 7.8 mJ $mol^{-1}$ $K^{-2}$ for 3R-$Ta_{0.9}Mo_{0.1}Se_2$. The normalized specific heat jump values $\Delta C/\gamma T_c$ are found to be 1.30 for 3R-$Ta_{0.9}W_{0.1}Se_2$ and 1.34 for 3R-$Ta_{0.9}Mo_{0.1}Se_2$, which are near the Bardeen-Cooper-Schrieffer (BCS) weak-coupling limit value (1.43), confirming bulk superconductivity. Using the Debye temperature ($\theta_D$), the critical temperature $T_c$, assuming a value[21] of μ* of 0.13 and that the electron-phonon coupling constant ($\lambda_{ep}$) can be calculated from the inverted McMillan formula [21]:

$$\lambda_{ep} = \frac{1.04 + \mu^* \ln\left(\frac{\theta_D}{1.45T_C}\right)}{(1 - 0.62\mu^*)\ln\left(\frac{\theta_D}{1.45T_C}\right) - 1.04}.$$

the values of $\lambda_{ep}$ obtained are 0.61 for 3R-$Ta_{0.9}W_{0.1}Se_2$ and 0.59 for 3R-$Ta_{0.9}Mo_{0.1}Se_2$ and suggest weak coupling superconductivity. With the Sommerfeld parameter ($\gamma$) and the electron-phonon coupling ($\lambda_{ep}$), the density of states (DOS) at the Fermi level can be calculated from $N(E_F) = \frac{3}{\pi^2 k_B^2 (1 + \lambda_{ep})}\gamma$. This yields $N(E_F)$ = 1.92 states/eV f.u. for optimal 3R-$Ta_{0.9}W_{0.1}Se_2$ and $N(E_F)$ = 1.56 states/eV f.u. for 3R-$Ta_{0.9}Mo_{0.1}Se_2$. The superconducting parameters of 3R-$Ta_{0.9}M_{0.1}Se_2$ (M = W, Mo) and $TaSe_{1.65}Te_{0.35}$ are summarized in Table 1, and can be compared to those 2H-$TaSe_2$.[22] Why these parameters are different in the 2H and 3R polytype is not currently known.

To gain further insight into the 3R materials, we now consider their calculated electronic band structures and densities of states (DOS). The calculated electronic structures are similar for both materials and so we focus on $Ta_{8/9}Mo_{1/9}Se_2$ ($Ta_{0.89}Mo_{0.11}Se_2$) in Figure 4a, which emphasizes contributions in the range around the Fermi level. Analysis of the orbitals contributing to the bands at $E_F$ shows the dominance of the transition metal states in this energy regime. Saddle points, which can be taken as an indication that the potential for superconductivity exists in a system [23], are seen at between F and G points in the Brillouin zone near the Fermi level. Figure 4a shows the calculated DOS for 3R-$Ta_{0.9}W_{0.1}Se_2$ and 3R-$Ta_{0.9}Mo_{0.1}Se_2$. The similarity in calculated electronic structures is readily seen, as is the fact that the filling has put the Fermi energy just above a peak in the DOS.

Finally, Figure 5 shows the variation of the superconducting $T_c$ with the structural character of the 2H, 3R and 1T superconducting forms of doped $TaSe_2$. The characteristic structural parameter is the *c/a* ratio of a single $MX_2$ layer plus its associated van der Waals gap, which we designate as the reduced *c*/*a* ratio (*c*/*n*)/*a*, where *n* is the number of layers in the unit cell stacking repeat. This is the classical way to characterize the anisotropy of hexagonal symmetry materials. The Figure compares the character of 3R-$Ta_{0.9}W_{0.1}Se_2$ and $Ta_{0.9}Mo_{0.1}Se_2$, with those of 2H-$TaSe_2$, 3R-$TaSe_{2-x}Te_x$ ($0.1 \leq x \leq 0.35$) and 1T-$TaSe_{2-x}Te_x$ ($0.8 \leq x \leq 1$) [17]. The $T_c$s for 3R-$Ta_{0.9}W_{0.1}Se_2$ and $Ta_{0.9}Mo_{0.1}Se_2$ are close to those of 3R-$TaSe_{1.65}Te_{0.35,}$ and the structural characteristics are comparable, even though the chemical systems themselves are very different.

## 4. Conclusion

The 3R-$Ta_{1-x}W_xSe_2$ ($0.07 \leq x \leq 0.125$) and 3R-$Ta_{1-x}Mo_xSe_2$ ($0.1 \leq x \leq 0.2$) have been synthesized successfully via a solid state reaction method. Based on the substitution of Mo for Ta in $TaSe_2$, the polytype found changed from 2H to 3R. 3R-$Ta_{0.9}W_{0.1}Se_2$ and 3R-$Ta_{0.9}Mo_{0.1}Se_2$ show a maximum $T_c \approx 2$ K, which is very similar to that of 3R-$TaSe_{1.65}Se_{0.35}$. The electronic structure calculations, through revealing a saddle point and peak in the density of states, confirm the electronic instability of the 3R-type system with respect to superconductivity. The results show that both isoelectronic X ion doping, through Te/Se substitution in 3R-$TaSe_{1.65}Se_{0.35,}$ and aliovalent M-site doping, through Mo or W substitution for Ta, lead to the stabilization of the 3R polytype and a dramatically increased $T_c$ over the more stable 2H polytype in the $TaSe_2$ system. Thus we conclude that rather than any special characteristics of the Te/Se substitution, it is the change from the 2H to the 3R polytype that results in the increase of $T_c$ by about an order of magnitude in the $TaSe_2$ system. The reason for this dramatic increase in $T_c$ is not yet known and its determination would be of interest for further study.

## Acknowledgements

The characterization of the electronic properties and the synthesis of the materials was supported by the DOE BES through grant DE-FG02-98ER45706, and the structural characterization and electronic structure calculations were supported by the Gordon and Betty Moore Foundation's EPiQS Initiative through Grant GBMF4412. The authors acknowledge Jason Krizan and Morten B. Nielsen for stimulating discussions.

**Table 1.** Superconducting parameters of 3R-$Ta_{0.9}M_{0.1}Se_2$ (M = W, Mo) and comparison to 3R-$TaSe_{1.65}Te_{0.35}$. Superconducting parameters of 3R-$TaSe_{1.65}Te_{0.35}$ were extracted from Reference 17. In all cases, a μ* of 0.13 has been assumed [21].

| Parameter | 3R-$Ta_{0.9}W_{0.1}Se_2$ | 3R-$Ta_{0.9}Mo_{0.1}Se_2$ | 3R-$TaSe_{1.65}Te_{0.35}$ |
|---|---|---|---|
| $T_c$ (K) | 2.2 | 2 | 2.4 |
| $a$ (Å) | 3.4222(2) | 3.4222(2) | 3.465(1) |
| $c$ (Å) | 19.1855(4) | 19.1734(5) | 19.59(6) |
| $\gamma$ (mJ mol$^{-1}$ K$^{-2}$) | 7.27 | 5.83 | 7.25 |
| $\Theta_D$ (K) | 203 | 198 | 184 |
| $\lambda_{ep}$ | 0.61 | 0.59 | 0.64 |
| $N(E_F)$ experiment (states/eV/f.u.) | 1.92 | 1.56 | 1.88 |
| $N(E_F)$ calculation (states/eV/f.u.) | 1.78 | 1.78 | - |
| $\Delta C/\gamma T_c$ | 1.31 | 1.34 | 1.2 |

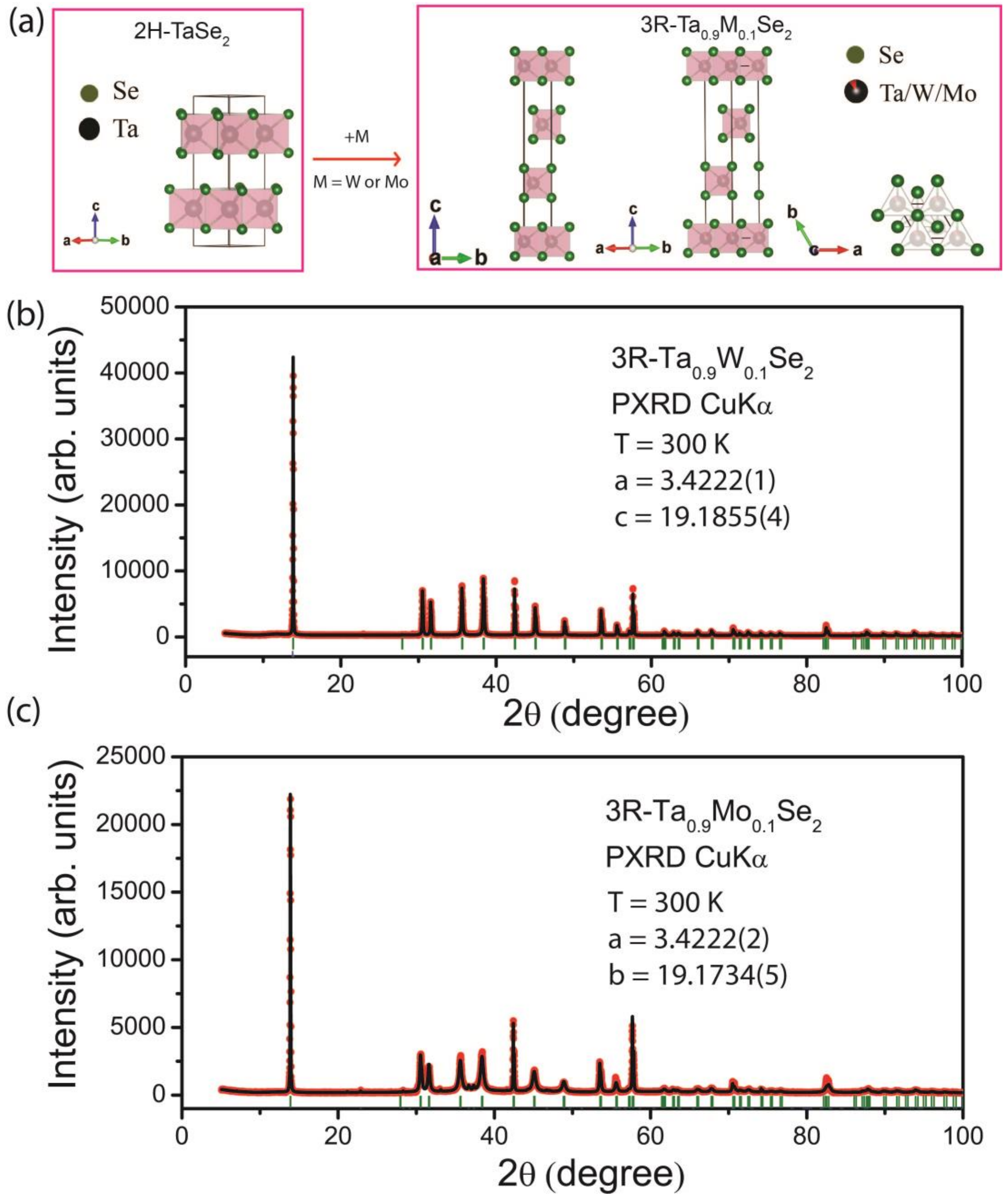

**Figure 1. Structural characterization of 3R-$Ta_{0.9}M_{0.1}Se_2$ (M = W, Mo).** (a) The crystal structures of 2H-$TaSe_2$ (left) and 3R-$Ta_{0.9}M_{0.1}Se_2$ (M = Mo, W) (right); (b) Powder X-ray diffraction pattern for 3R-$Ta_{0.9}W_{0.1}Se_2$; (c) Powder X-ray diffraction pattern for 3R-$Ta_{0.9}Mo_{0.1}Se_2$. Green tics show calculated peak positions for the 3R phase.

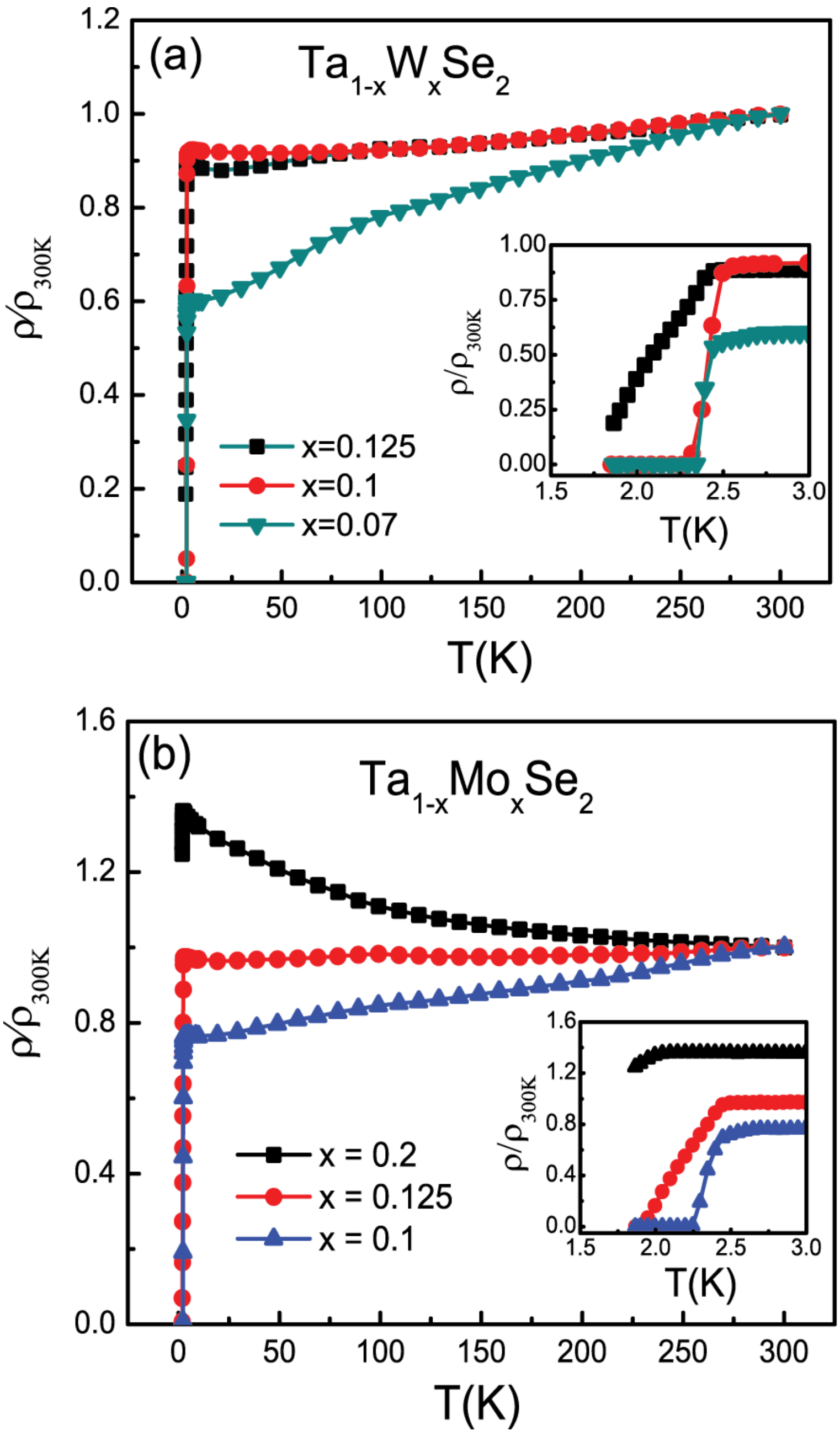

**Figure 2**. **Transport characterization of the normal states and superconducting transitions.** (a) The temperature dependence of the ratio ($\rho/\rho_{300K}$) for 3R-$Ta_{1-x}W_xSe_2$ ($0.07 \leq x \leq 0.125$). Inset: enlarged view of low temperature region (1.5 - 3 K), showing the superconducting transition. (b) The temperature dependence of the ratio ($\rho/\rho_{300K}$) for 3R-$Ta_{1-x}Mo_xSe_2$ ($0.1 \leq x \leq 0.2$), Inset: enlarged view of low temperature region (1.5 - 3 K) showing the superconducting transition.

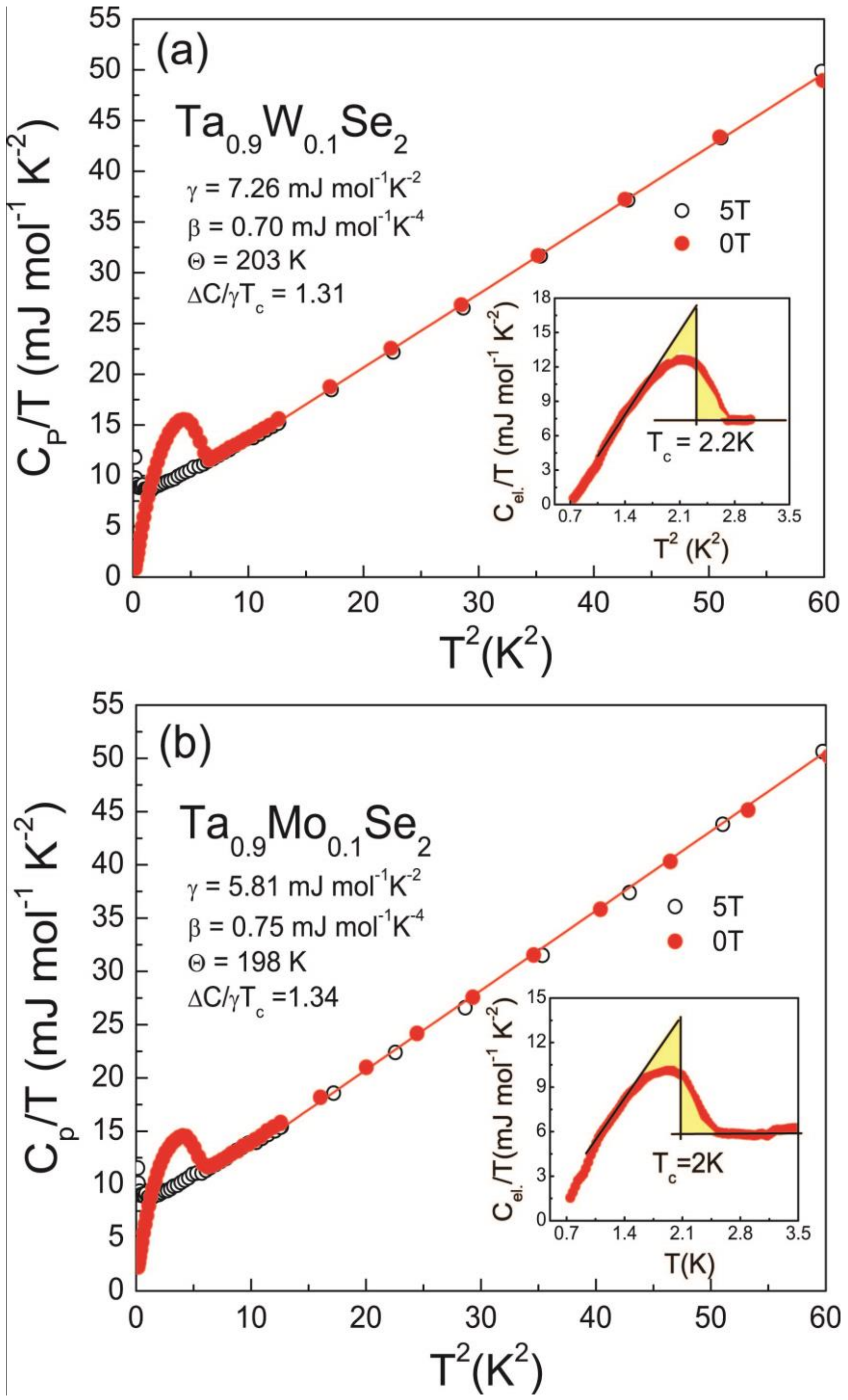

**Figure 3**. **Specific heat characterization of the superconductivity** (a) The temperature dependence of the specific heat $C_p$ of 3R-$Ta_{0.9}W_{0.1}Se_2$, presented in the form of $C_p/T$ vs. $T^2$ (main panel) and $C_{el}/T$ vs. T (inset). (b) The temperature dependence of specific heat $C_p$ of 3R-$Ta_{0.9}Mo_{0.1}Se_2$, presented in the form of $C_p/T$ vs $T^2$ (main panel) and $C_{el}/T$ vs T (inset). Yellow regions in the insets show the equal area constructions.

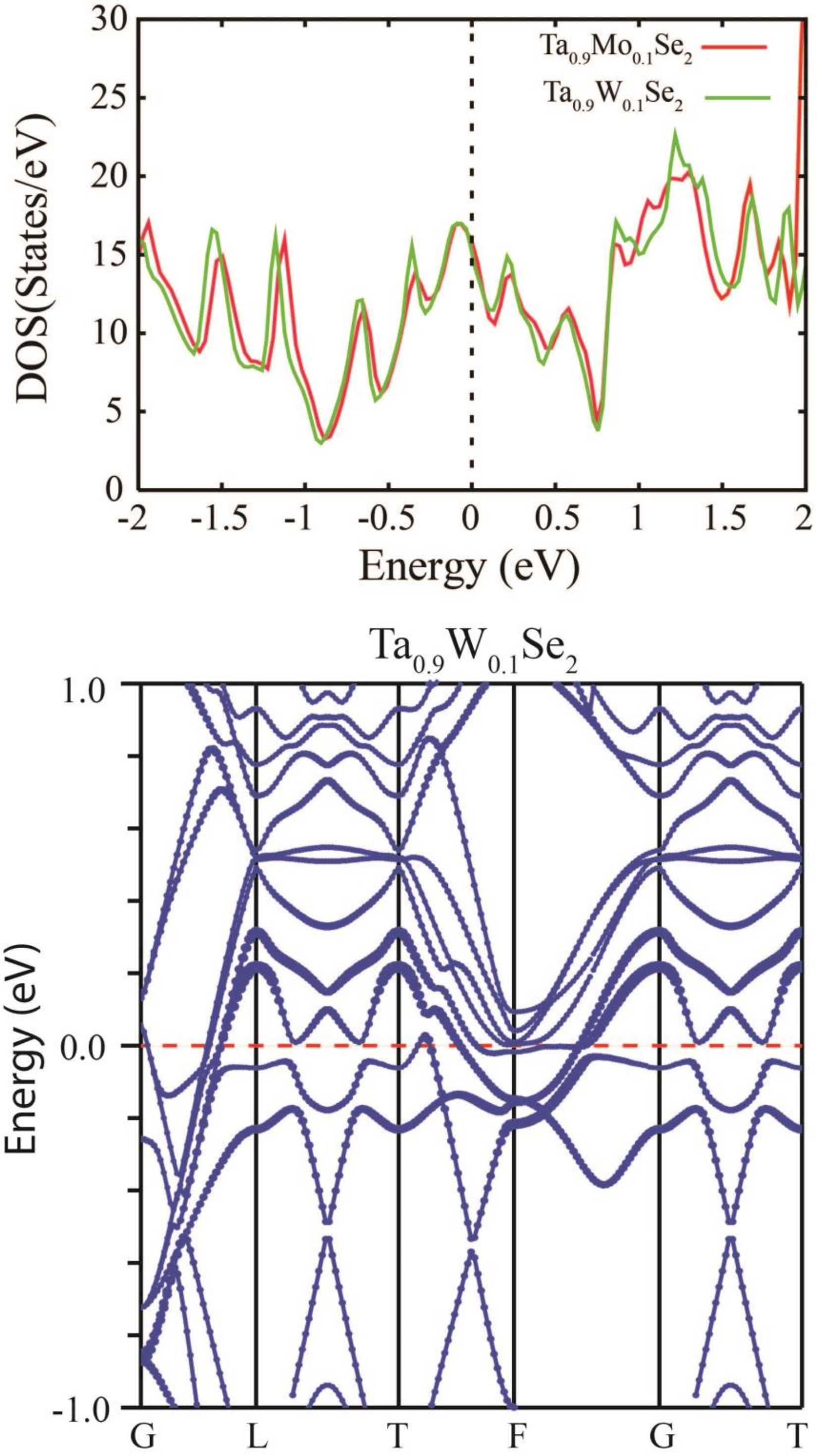

**Figure 4. The calculated electronic structures of the superconducting 3R phases.** (a) The calculated density of states (DOS) for 3R $Ta_{0.9}Mo_{0.1}Se_2$ and $Ta_{0.9}W_{0.1}Se_2$. (b) The calculated band structure for $Ta_{0.9}W_{0.1}Se_2$. Total DOS curves and band structure curves are obtained from non-spin-polarized LDA calculations.

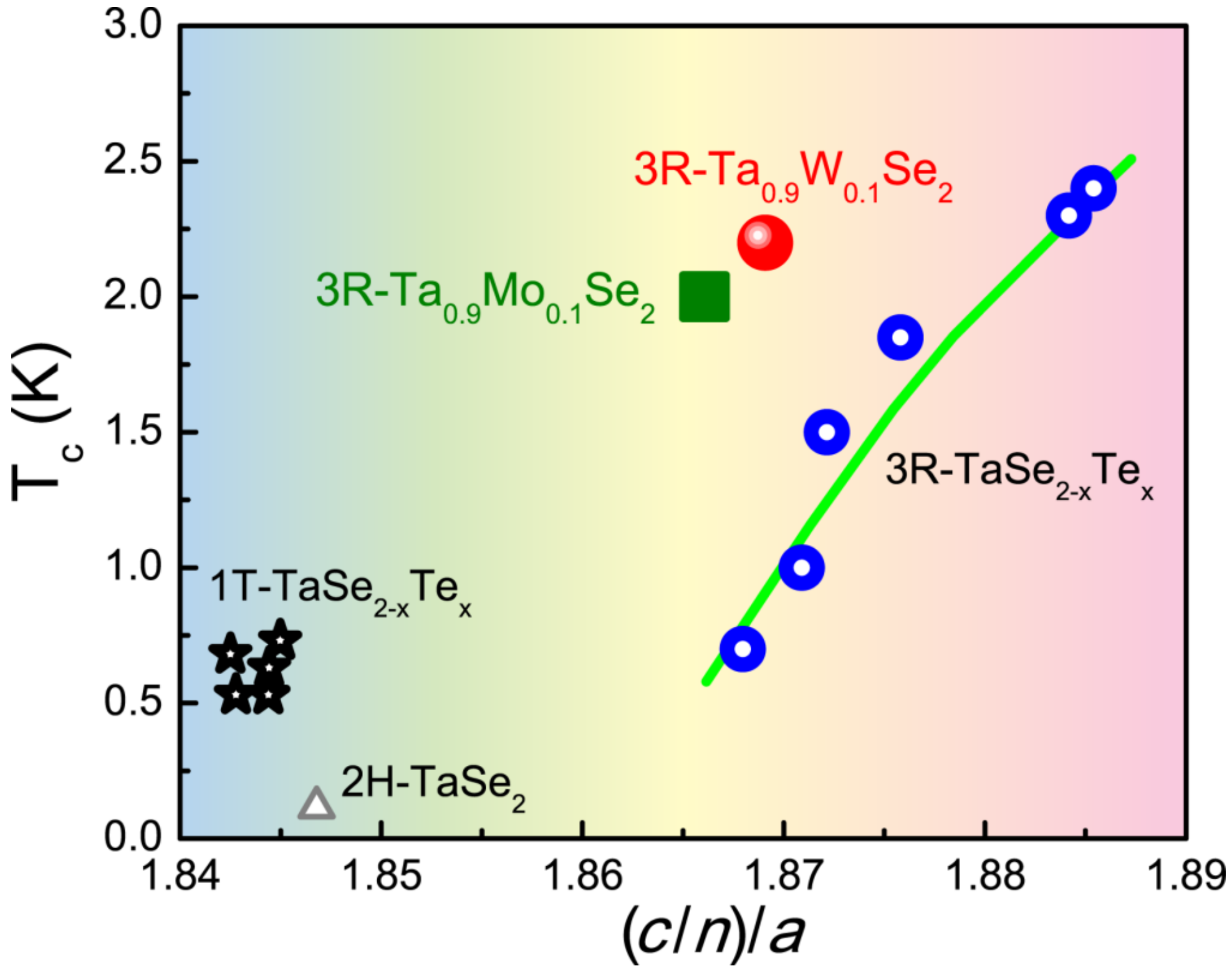

**Figure 5**. **Structure-$T_c$ correlations in the $TaSe_2$ system.** The variation of the superconducting $T_c$ with the reduced *c*/*a* ratio, (*c*/*n*)/*a*, for 3R-$Ta_{0.9}W_{0.1}Se_2$ and $Ta_{0.9}Mo_{0.1}Se_2$, comparison with 2H-$TaSe_2$, 1T-$TaSe_{2-x}Te_x$ ($0.8 \leq x \leq 1$) and 3R-$TaSe_{2-x}Te_x$ ($0.1 \leq x \leq 0.35$), (the data for $TaSe_{2-x}Te_x$ is from Reference 17), where *n* = number of layers in the stacking repeat, and *c* and *a* are the unit cell parameters.